\documentclass[vecphys]{svmult}
\usepackage{makeidx}
\usepackage{graphicx}
\usepackage{multicol}
\usepackage{amssymb}
\usepackage{cite}
\usepackage[bottom]{footmisc}
\usepackage{epstopdf}
\makeindex
\begin{document}

\title{Spin-lattice coupling in frustrated antiferromagnets}
\titlerunning{Spin-lattice coupling}
\author{Oleg Tchernyshyov$^1$ \and Gia-Wei Chern$^2$}
\authorrunning{O. Tchernyshyov \and G.-W. Chern}
\institute{
Department of Physics and Astronomy,
Johns Hopkins University,
3400 N. Charles St.,
Baltimore, MD 21218, U. S. A.
\and 
Department of Physics, 
University of Wisconsin, 
1150 University Avenue, 
Madison, WI 53706, U. S. A.
}

\maketitle

We review the mechanism of spin-lattice coupling in relieving the
geometrical frustration of pyrochlore antiferromagnets, in
particular spinel oxides. The tetrahedral unit, which is the
building block of the pyrochlore lattice, undergoes a spin-driven
Jahn-Teller instability when lattice degrees of freedom are coupled to 
the antiferromagnetism. By restricting our considerations to distortions 
which preserve the translational symmetries of the lattice, we present a 
general theory of the collective spin-Jahn-Teller effect in the pyrochlore 
lattice. One of the predicted lattice distortions breaks the inversion 
symmetry and gives rise to a chiral pyrochlore lattice, in which frustrated
bonds form helices with a definite handedness. The chirality is transferred 
to the spin system through spin-orbit coupling, resulting in a long-period 
spiral state, as observed in spinel CdCr$_2$O$_4$. We discuss explicit 
models of spin-lattice coupling using local phonon modes, and their 
applications in other frustrated magnets.

\section{Introduction}
\label{sec:intro}

As explained in the introductory chapter by Chalker, sufficiently 
strong frustration in a magnet results in a large degeneracy of its 
ground-state manifold. Prime examples of this behavior are Heisenberg
antiferromagnets on the kagome \cite{Ramirez91JAP,Huse92PRB} and
pyrochlore \cite{Zinkin94PRL, Moessner98PRB} lattices with
interactions restricted to nearest-neighbor sites. In the classical
limit of a large spin $S$, the ground states of these magnets exhibit 
very high, continuous degeneracies and possess numerous zero modes, 
which correspond to moving the system through its manifold of ground 
states \cite{Chalker92PRL}. The pyrochlore antiferromagnet represents 
a particularly striking example of high ground-state degeneracy: at
least half of its spin-wave modes have zero frequencies in any
collinear ground state \cite{Moessner98PRB}.

A large degeneracy means enhanced sensitivity to perturbations, even
when these are nominally weak. In this chapter we will consider 
a coupling between spins and the underlying lattice, which has its 
origin in the dependence of the exchange integrals on the atomic
positions, $J(\mathbf r_1, \mathbf r_2) \, \mathbf S_1 \cdot \mathbf S_2$,
and is known as magnetoelastic exchange \cite{Kittel60}. In the pyrochlore
antiferromagnet, this coupling lifts the degeneracy of the classical
ground states and induces a symmetry-lowering distortion of the lattice,
in analogy with the spin-Peierls effect in antiferromagnetic spin chains
\cite{PhysRevB.14.3036}. A spin-Peierls-like phase transition has
been observed in several antiferromagnetic spinels where the magnetic 
ions form the pyrochlore lattice \cite{PhysRevLett.84.3718,
chung:247204, ueda:094415}.

The problem of coupled spins and lattice degrees of freedom in a pyrochlore
antiferromagnet is reminiscent of the collective Jahn-Teller effect
\cite{Bersuker} in crystalline solids. We therefore begin the discussion 
by studying the Jahn-Teller distortion in a tetrahedral ``molecule'' with 
four spins, which is the structural unit of the pyrochlore lattice. A 
symmetry-based analysis will be supplemented by models with specific 
spin-phonon coupling mechanisms. We will then extend the analysis to 
an infinite lattice, to examine some of the possible ground states of 
the classical spin system. In concluding this chapter, we will test the 
theory of spin-phonon coupling on the example of CdCr$_2$O$_4$, a 
frustrated Heisenberg antiferromagnet with $S = 3/2$ spins residing on 
the pyrochlore lattice.

\section{Spin-driven Jahn-Teller effect in a tetrahedron}
\label{sec:JT}

\begin{figure}
\begin{center}
\includegraphics[width=0.3\columnwidth]{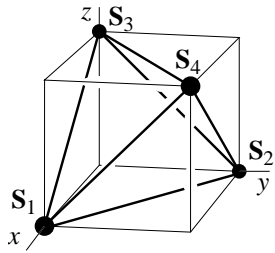}
\end{center}
\caption{Representation of four spins in the corners of a regular 
tetrahedron.}
\label{fig-tetrahedron}
\end{figure}

Consider the basic structural unit of the pyrochlore lattice, a cluster in 
the shape of a regular tetrahedron with four spins of length $S$ at the
corners (Fig.~\ref{fig-tetrahedron}). The Heisenberg exchange energy
depends on the total spin of the cluster, $\mathbf S_\mathrm{tot} = 
\sum_{i=1}^4 \mathbf S_i$, according to
\begin{equation}
H_0 = J \sum_{i<j} \mathbf S_i \cdot \mathbf S_j
= \frac{J \mathbf S_\mathrm{tot}^2}{2}  - \mathrm{const}.
\label{eq:H0}
\end{equation}
For antiferromagnetic exchange ($J>0$), this energy is minimized when
the total spin of the cluster is 0. The degeneracy of the ground state
is thus equal to the number of distinct spin-singlet states in a system
of four spins.

For spins of length $S$, the number of linearly independent singlet
ground states is $2S + 1$.  Indeed, the pair of spins 1 and 2 can have a 
total combined spin $\mathbf S_{12}$ ranging from 0 to $2S$.  The same is 
true of the spin pair 3 and 4. A state with a total spin $S_\mathrm{tot}
 = 0$ can be obtained by combining states with $S_{12} = S_{34} = 0$, 
$S_{12} = S_{34} = 1$, and so on, up to $S_{12} = S_{34} = 2S$. Thus 
one observes that, for any spin length $S$, the ground state of the four
spins is degenerate.  The high symmetry of the cluster means that,
in accordance with the Jahn-Teller theorem,\footnote{It is important to
note that the degeneracy is not caused by the symmetry of time reversal,
so the conditions of the theorem are fulfilled \cite{Jahn37}.} the
ground-state energy can be lowered through a distortion.

Assume for simplicity (a more general case will be considered below) 
that the exchange coupling $J$ between spins $i$ and $j$ has a dependence 
on their separation, $r_{ij} = |\mathbf r_i - \mathbf r_j|$, given by 
\begin{equation}
J(r_{ij}) = J(\bar{r}_{ij}) + J'(\bar{r}_{ij}) \delta r_{ij}
+ J''(\bar{r}_{ij}) \delta r_{ij}^2/2 + \ldots,
\label{eq:J-of-r}
\end{equation}
where $\bar{r}_{ij}$ is a reference distance. At zeroth order in the 
displacements $\delta \mathbf r_i$, we recover the unperturbed Heisenberg 
Hamiltonian (\ref{eq:H0}) with $2S + 1$ degenerate singlet ground states.

The first-order term,
\begin{equation}
H_1 = J' \sum_{i<j} (\mathbf S_i \cdot \mathbf S_j)
\delta r_{ij}, \label{eq:H1}
\end{equation}
lifts the degeneracy of the ground-state manifold. As long as the 
displacements involved remain small enough to satisfy $J' \delta r_{ij} 
\ll J$, excited states with $S_\mathrm{tot} > 0$ at energies $J$ and 
higher may be neglected. Thus it is necessary to determine the energy 
levels of $H_1$ in the Hilbert space of the singlet ground states.

\subsection{Generalized coordinates and forces}
\label{sec:JT-coords}

The perturbation Hamiltonian (\ref{eq:H1}) depends on the atomic
displacements $\delta \mathbf r_i$, which comprise $4 \times 3 = 12$
degrees of freedom. However, not all of these will influence the
exchange energy of the spins. As one example, this does not change under
rigid translations of the tetrahedron (3 modes) or under global rotations
(a further 3 modes). The remaining 6 modes represent various deformations 
of the four-site cluster. To facilitate further analysis, we classify these 
modes in terms of the irreducible representations (irreps) of the tetrahedral
point group $T_d$ \cite{LL}.

The 6 modes belong to three irreps of $T_d$. The breathing mode
(irrep $A$) leaves the symmetry of the tetrahedron fully intact. A
doublet of tetragonal and orthorhombic distortions, $\mathbf Q^E =
(Q^E_1, Q^E_2)$, transforms under irrep $E$.  Finally, a triplet
$\mathbf Q^{T_2} = (Q^{T_2}_1, Q^{T_2}_2, Q^{T_2}_3)$, transforming
as irrep $T_2$, elongates and compresses opposing bonds; equal-amplitude
superpositions of the triplet components yield trigonal distortions.
The coordinates of these modes can be expressed in terms of Cartesian 
displacements of the spins with the coefficients listed in
Table \ref{tab-coeff}.

\def\1{\frac{1}{\sqrt{12}}}
\def\2{\frac{1}{\sqrt{8}}}
\def\3{\frac{1}{\sqrt{24}}}
\def\4{\frac{1}{\sqrt{6}}}

\begin{table}
\caption{Coefficients relating the 6 distortions of a tetrahedron to the
displacements $\delta \mathbf r_i$ of its vertices.  The reference frame
is shown in Fig.~\ref{fig-tetrahedron}.}
\label{tab-coeff}
\begin{tabular}{|l|c|ccc|ccc|ccc|ccc|}
\hline
 &
 & $\delta x_1$ & $\delta y_1$ & $\delta z_1$
 & $\delta x_2$ & $\delta y_2$ & $\delta z_2$
 & $\delta x_3$ & $\delta y_3$ & $\delta z_3$
 & $\delta x_4$ & $\delta y_4$ & $\delta z_4$\\
 \hline
$A$ & $Q^A$
    & $+\1$ & $-\1$ & $-\1$
    & $-\1$ & $+\1$ & $-\1$
    & $-\1$ & $-\1$ & $+\1$
    & $+\1$ & $+\1$ & $+\1$\\
\hline
$E$ & $Q^E_1$
    & $-\3$ & $+\3$ & $-\4$
    & $+\3$ & $-\3$ & $-\4$
    & $+\3$ & $+\3$ & $+\4$
    & $-\3$ & $-\3$ & $+\4$ \\
    & $Q^E_2$
    & $+\2$ & $+\2$ & 0
    & $-\2$ & $-\2$ & 0
    & $-\2$ & $+\2$ & 0
    & $+\2$ & $-\2$ & 0 \\
\hline
$T_2$& $Q^{T_2}_1$
    & 0 & $+\2$ & $+\2$
    & 0 & $+\2$ & $-\2$
    & 0 & $-\2$ & $+\2$
    & 0 & $-\2$ & $-\2$ \\
    & $Q^{T_2}_2$
    & $+\2$ & 0 & $-\2$
    & $+\2$ & 0 & $+\2$
    & $-\2$ & 0 & $+\2$
    & $-\2$ & 0 & $-\2$ \\
    & $Q^{T_2}_3$
    & $+\2$ & $-\2$ & 0
    & $-\2$ & $+\2$ & 0
    & $+\2$ & $+\2$ & 0
    & $-\2$ & $-\2$ & 0 \\
\hline
\end{tabular}
\end{table}

\def\5{\sqrt{\frac{2}{3}}}
\def\6{\frac{1}{\sqrt{3}}}

\begin{table}
\caption{Coefficients relating the bond elongations $\delta r_{ij}$ to
the distortion coordinates $Q^A$, $\mathbf Q^E = (Q^E_1, Q^E_2)$, and
$\mathbf Q^{T_2} = (Q^{T_2}_1, Q^{T_2}_2, Q^{T_2}_3)$.}
\begin{tabular}{|l|c|cc|ccc|}
\hline
 & $Q^A$ & $Q^E_1$ & $Q^E_2$ & $Q^{T_2}_1$ & $Q^{T_2}_2$ & $Q^{T_2}_3$\\
 \hline
$\delta r_{14}$ & $+\5$ & $+\1$ & $-\frac{1}{2}$ & $-1$ & 0 & 0 \\
$\delta r_{23}$ & $+\5$ & $+\1$ & $-\frac{1}{2}$ & $+1$ & 0 & 0 \\
\hline
$\delta r_{24}$ & $+\5$ & $+\1$ & $+\frac{1}{2}$ & 0 & $-1$ & 0 \\
$\delta r_{13}$ & $+\5$ & $+\1$ & $+\frac{1}{2}$ & 0 & $+1$ & 0 \\
\hline
$\delta r_{34}$ & $+\5$ & $-\6$ & 0 & 0 & 0 & $-1$ \\
$\delta r_{12}$ & $+\5$ & $-\6$ & 0 & 0 & 0 & $+1$ \\
\hline
\hline
\end{tabular}
\end{table}

By expressing the changes in bond lengths, $\delta r_{ij}$, in terms of 
the generalized coordinates, we reduce the perturbation Hamiltonian
(\ref{eq:H1}) to the form
\begin{equation}
H_1 = \sum_{\alpha} {J^\alpha}' \sum_{n} Q^\alpha_n f^\alpha_n,
\end{equation}
where the index $\alpha$ enumerates the irreps and $n$ its components. 
The variable $f^\alpha_n$ is the generalized force which is conjugate to 
the coordinate $Q^\alpha_n$ and has the same symmetry properties. 
${J^\alpha}'$ is a coupling constant appropriate for the irrep $\alpha$.

For illustration, the breathing mode $Q_A$ couples to the spin operator
\begin{equation}
f^A = \frac{1}{\sqrt{6}} \sum_{i<j} \mathbf S_i \cdot \mathbf S_j,
\end{equation}
which is invariant under all symmetry operations of $T_d$. Further,
up to a trivial factor, this is the unperturbed Hamiltonian (\ref{eq:H0}) 
and so has the same value in any of the degenerate ground states. 
Consequently, the term $-{J^A}' Q^A f^A$ in the perturbation Hamiltonian 
(\ref{eq:H1}) produces a trivial energy shift of the degenerate ground 
states, but does not split them.

The triplet mode and the associated triplet force also do not induce a
splitting.  To demonstrate this, we note that $Q^{T_2}_1$ couples to the
operator
\begin{equation}
f^{T_2}_1 = (\mathbf S_2 \cdot \mathbf S_3 - \mathbf S_1 \cdot \mathbf S_4)
/\sqrt{2},
\end{equation}
which vanishes in any singlet state.\footnote{Indeed, from $\mathbf
S_1 + \mathbf S_2 + \mathbf S_3 + \mathbf S_4 = 0$ it follows that
$(\mathbf S_1 + \mathbf S_4)^2 = (\mathbf S_2 + \mathbf S_3)^2$ and
thus $\mathbf S_1 \cdot \mathbf S_4 = \mathbf S_2 \cdot \mathbf
S_3$.}  In the presence of an applied magnetic field, the triplet forces 
cannot be neglected because of the nonzero total spin $S_{\rm tot}$. The
triplet forces and the corresponding trigonal distortions play an
important role in the stabilization of the half-magnetization plateaus 
observed in some spinel chromites (Sec.~\ref{sec:plateau}).

The only two modes involved in the splitting of the ground state are the
components of the doublet $(Q^E_1, Q^E_2)$ of tetragonal and orthorhombic 
distortions. These couple, respectively, to the spin operators
\begin{eqnarray}
f^E_1 &=&
    \frac{\mathbf S_1 \cdot \mathbf S_4
        + \mathbf S_2 \cdot \mathbf S_3
        + \mathbf S_2 \cdot \mathbf S_4
        + \mathbf S_1 \cdot \mathbf S_3
       -2 \mathbf S_1 \cdot \mathbf S_2
       -2 \mathbf S_3 \cdot \mathbf S_4}{\sqrt{12}},
\nonumber
\\
f^E_2 &=&
    \frac{\mathbf S_2 \cdot \mathbf S_4
        + \mathbf S_1 \cdot \mathbf S_3
        - \mathbf S_1 \cdot \mathbf S_4
        - \mathbf S_2 \cdot \mathbf S_3
    }{2}.
\label{eq:f}
\end{eqnarray}
In what follows, we omit the irrep superscript because only the doublet 
$E$ participates in lifting the degeneracy of the ground-state manifold.

In addition to the magnetoelastic exchange (\ref{eq:H1}), which is linear
in the distortion amplitude, it is necessary to consider also the elastic 
energy of the deformation. We thus obtain the spin-distortion Hamiltonian
\begin{equation}
H = J'(\mathbf Q \cdot \mathbf f) + k |\mathbf Q|^2/2
\equiv J'(Q_1 f_1 + Q_2 f_2) + k(Q_1^2 + Q^2_2)/2,
\label{eq:H}
\end{equation}
where $k$ is the elastic constant of the doublet distortion. Having 
established this, the next task is to minimize the energy (\ref{eq:H}) 
with respect to both the coordinates and the spins.

\subsection{Four $S = 1/2$ spins on a tetrahedron}
\label{sec:JT-half}

The problem of four $S = 1/2$ spins on a deformable tetrahedron was
first analyzed by Yamashita and Ueda \cite{PhysRevLett.85.4960}.
The ground state of the unperturbed exchange Hamiltonian
(\ref{eq:H0}) is two-fold degenerate. As a basis in this Hilbert
space one may use singlet states with a well-defined total spin on bonds
12 and 34, $S_{12} = S_{34} = \sigma$, where $\sigma = 0$ or 1
\cite{Tch02PRB}. In this basis, the force operators $f_1$ and $f_2$ are
proportional to the Pauli matrices $\sigma_1$ and $\sigma_3$, respectively,
so that the Hamiltonian (\ref{eq:H}) reduces to
\begin{equation}
H = J'(Q_1 \sigma_1 + Q_2 \sigma_3) \sqrt{3}/2 + k(Q_1^2 + Q^2_2)/2.
\label{eq:H-half}
\end{equation}

For a given distortion $\mathbf Q$, the ground-state manifold is split
into the two energy levels
\begin{equation}
E_{1,2} = \pm |J'|Q\sqrt{3}/2 + kQ^2/2,
\end{equation}
and the energy of the system is minimized when $Q = |J'|\sqrt{3}/(2k)$.
Note that this quantity depends on the magnitude of the distortion,
$Q = \sqrt{Q_1^2 + Q_2^2}$, but not on its ``direction:''
it can be tetragonal, purely orthorhombic, or any combination of these. 
This degeneracy is associated with a continuous symmetry of the Hamiltonian
(\ref{eq:H-half}) that involves simultaneous ``rotations'' of both the
distortion coordinates and the Pauli matrices,
\begin{equation}
 \left(
  \begin{array}{l}
   Q_1 \\ Q_2
  \end{array}
 \right)
 \mapsto
 \left(
  \begin{array}{rr}
   \cos{\theta} & \sin{\theta}\\
   -\sin{\theta} & \cos{\theta}
  \end{array}
 \right)
 \left(
  \begin{array}{l}
   Q_1 \\ Q_2
  \end{array}
 \right),
\quad
 \left(
  \begin{array}{l}
   \sigma_1 \\ \sigma_3
  \end{array}
 \right)
 \mapsto
 \left(
  \begin{array}{rr}
   \cos{\theta} & \sin{\theta}\\
   -\sin{\theta} & \cos{\theta}
  \end{array}
 \right)
 \left(
  \begin{array}{l}
   \sigma_1 \\ \sigma_3
  \end{array}
 \right).
\end{equation}
The invariance of the energy under this transformation does not reflect
an underlying symmetry, and applies at the level of approximation used 
here.  Terms of higher order in $\mathbf Q$ break this symmetry to leave 
only a three-fold degeneracy, as may be expected on symmetry grounds
\cite{PhysRevLett.85.4960}.  The lowest-order anharmonic term allowed by the
symmetry is proportional to
\begin{equation}
Q_x Q_y Q_z \equiv
 \left(
  -\frac{1}{2}Q_1 + \frac{\sqrt{3}}{2}Q_2
 \right)
 \left(
  -\frac{1}{2}Q_1 - \frac{\sqrt{3}}{2}Q_2
 \right)
 Q_1 =
\frac{1}{4}Q^3\cos{3 \alpha},
\end{equation}
where $Q_x$, $Q_y$, and $Q_z$ measure tetragonal distortions along the
respective axes and $\alpha$ is the polar angle in the $(Q_1,Q_2)$ plane.
Depending on the sign of the cubic term, it favors three distinct ground 
states at $\alpha = 0, \pm 2\pi/3$ or at $\alpha = \pi, \pm \pi/3$. The 
former ``vacua'' have spin singlets on two opposing bonds 
(\textit{e.g.}~$S_{12} = S_{34} = 0$ for $\alpha = 0$) while 
the latter have spin triplets on two opposing bonds ($S_{12} = S_{34} = 1$
for $\alpha = \pi$). These ground states exhibit valence-bond order,
which violates the point-group symmetry of the cluster but not the
SO(3) symmetry of the exchange interaction. The two-component
order parameter (\ref{eq:f}), introduced by Harris, Berlinsky and
Bruder \cite{harris:5200}, measures the differences in spin
correlations on the different bonds.

The ground states of the cluster exhibit a tetragonal lattice distortion
along one of the three major axes with
$\mathbf Q = -J' \mathbf \langle \mathbf f \rangle \sqrt{3}/(2k)$.
If $J'<0$, as would be expected for direct antiferromagnetic exchange,
the tetrahedron is flattened (elongated) in a ground state with triplets
(singlets) on opposing bonds.

For spins of length $S > 1/2$, the analysis proceeds by a similar route
\cite{Tch02PRB}. The lowest-order perturbation (\ref{eq:H1}) yields three 
degenerate singlet ground states with the largest possible spins on two 
opposing bonds, such as $S_{12} = S_{34} = 2S$, and a tetragonal distortion 
(a flattening of the tetrahedron for $J' < 0$).  This result is most easily 
understood in the classical limit $S \to \infty$, to which we turn next.

\subsection{Four classical spins on a tetrahedron}
\label{sec:JT-classical}

For classical spins, the total energy $E(\mathbf {f,Q})$ (\ref{eq:H}) 
can be minimized in two steps. We minimize it first with respect to the
distortion $\mathbf Q$ at a fixed $\mathbf f$.\footnote{This method cannot
be applied to quantum spins because the operators $f_1$ and $f_2$ do not 
commute \cite{Tch02PRB}, and so their values cannot be measured 
simultaneously.} A minimum is achieved when $\mathbf Q = -J' \mathbf f/k$, 
yielding the energy 
\begin{equation}
E(\mathbf f) = -{J'}^2 f^2/(2k),
\label{eq:E-cl}
\end{equation}
whence the total energy is minimized by states with the largest magnitude 
of the force doublet $\mathbf f$. Thus it is necessary to quantify the 
magnetoelastic forces in those states of the ground-state manifold with 
$S_\mathrm{tot} = 0$.

\begin{figure}
\begin{center}
\includegraphics[width=0.8\columnwidth]{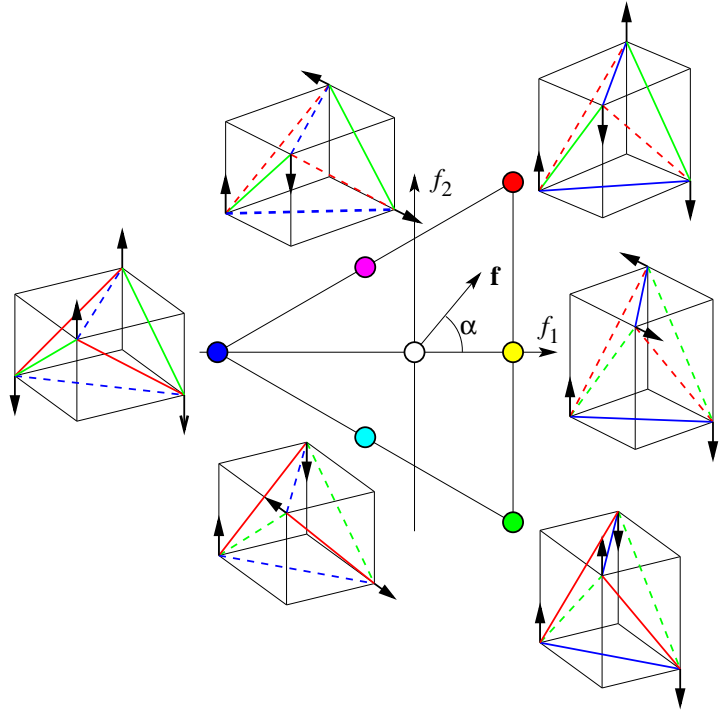}
\end{center}
\caption{The domain of attainable values of the force doublet
$\mathbf f = (f_1, f_2)$ (\ref{eq:f}) for classical spins.  Dashed lines
indicate frustrated bonds ($\mathbf S_i \cdot \mathbf S_j \geq 0$).  
Reprinted with permission from Ref.~\cite{Tch02PRB}.}
\label{fig:f}
\end{figure}

The domain of attainable $\mathbf f$ values forms a regular triangle
in the $(f_1, f_2)$ plane (Fig.~\ref{fig:f}).  Its three corners
correspond to the three distinct collinear states with four satisfied
bonds ($\mathbf S_i \cdot \mathbf S_j = - S^2$) and two frustrated ones
($\mathbf S_i \cdot \mathbf S_j = + S^2$). States elsewhere on the 
perimeter of the triangle have coplanar spins.

Not unexpectedly, the doublet force is maximized (and the total energy
minimized) in the collinear states. Indeed, in such states antiparallel
spins attract and parallel spins repel each other with forces
$F = -J' \, \mathbf S_i \cdot \mathbf S_j$ of the largest possible
magnitude, $|J'|S^2$. The large forces result in large distortions, 
yielding a large decrease in the total energy. Thus in the classical
limit one expects collinear ground states in which the tetrahedron is 
flattened along one of the principal axes for $J' < 0$.\footnote{If 
$J' > 0$, the spins are still collinear but the tetrahedron is 
elongated along the same axis.}

\subsection{Color notation and other useful analogies}

We find it convenient here to introduce an analogy with the color triangle, 
where the vertices correspond to the primary colors red, green, and blue,
the midpoints of the edges to the secondary colors cyan, magenta, and yellow,
and the center to the absence of color.  If we color bonds perpendicular to
the major axes $a$, $b$, and $c$ respectively red, green, and blue, then
the color of the state in Fig.~\ref{fig:f} reflects the color of the
frustrated bonds.

The collinear ground states obtained for classical spins provide a simple 
rationalization for the valence-bond states found for quantum spins of 
length $S > 1/2$ \cite{Tch02PRB}. Quantum states maximizing the spins of 
opposing pairs (for example $S_{12} = S_{34} = 2S$) are the analogs of 
the collinear classical configurations ($S_1 = S_2 = -S_3 = -S_4$).

Lastly, we recall that the spin-distortion Hamiltonian (\ref{eq:H}) was
derived for a simplified model of exchange in which the Heisenberg
interaction strength $J$ depends only on the separation of the two spins.
This approximation is good for direct spin exchange \cite{motida:1970}, 
which is the dominant exchange interaction in the chromium spinels 
ZnCr$_2$O$_4$ \cite{sushkov:137202} and CdCr$_2$O$_4$ \cite{aguilar:092412}, 
as well as in some other chromium antiferromagnets \cite{olariu:167203}. 
However, in other situations $J$ may exhibit a more complex dependence on 
atomic displacements, such as the very sensitive bond-angle-dependence of 
superexchange. Fortunately, the form of the spin-distortion coupling 
derived above (\ref{eq:H}) is robust, as can be seen by symmetry
considerations: group theory guarantees that there are no other invariant
terms of the same order in $\mathbf f$ and $\mathbf Q$. In the general 
case, $J'$ represents a linear combination of exchange derivatives.

\subsection{Spin-Jahn-Teller effect on a triangle}

Another well-known lattice producing strong frustration in an
antiferromagnet is the kagome geometry \cite{Ramirez91JAP,Huse92PRB}, 
a network of corner-sharing triangles in a plane, and its three-dimensional 
variant, the hyperkagome lattice \cite{PhysRevLett.99.137207}. It is natural 
to ask whether spin-lattice coupling is also an effective mechanism for 
relieving frustration in this case.  We answer this question by considering 
the building block of the kagome lattice, a triangle with three spins.

Classical Heisenberg spins interacting through antiferromagnetic
Heisenberg exchange minimize their energy by making angles of $120^\circ$
with one another. An analysis along the lines of Sec.~\ref{sec:JT}
shows that the correction to the exchange energy from the magnetoelastic
term is quadratic in the spin displacements: the linear term cancels.
Without this linear term, a spontaneous distortion does not occur. The
absence of the linear term can be understood simply from the
magnetoelastic forces between the spins: the three forces, being 
proportional to the scalar products of the spins, are equal in a state 
where the spins make equal angles with each other. These symmetrical 
forces only shrink the triangle without distorting it.

The argument against the Jahn-Teller distortion fails if the quadratic term
in the magnetoelastic correction is negative and large enough to overcome
the purely elastic cost of the distortion, a scenario proposed recently 
by Wang and Vishwanath \cite{wang:08prl}. In our view, however, empirical 
evidence indicates that a Jahn-Teller instability of this sort would be a 
rare exception. The strength of the quadratic magnetoelastic term can be 
estimated from the splitting of degenerate phonons in antiferromagnets with 
a spin-induced Jahn-Teller distortion \cite{sushkov:137202}. Such splittings, 
observed in ZnCr$_2$O$_4$ \cite{sushkov:137202}, CdCr$_2$O$_4$ 
\cite{aguilar:092412}, and MnO \cite{rudolf:024421}, do not exceed 
12\% of the phonon frequencies, which indicates the dominance of the 
purely elastic contribution.

\begin{figure}
\begin{center}
\includegraphics[width=0.4\columnwidth]{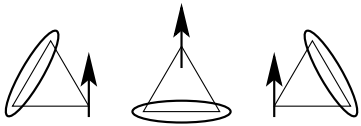}
\end{center}
\caption{Valence-bond ground states of three spins $S = 1/2$ with 
antiferromagnetic Heisenberg exchange interactions. The ellipses 
designate singlet bonds.}
\label{fig:vbs-triangle}
\end{figure}

While the spin-Jahn-Teller distortion of a triangle appears unlikely
in the classical limit, the opposite extreme --- quantum spins of length
$S = 1/2$ --- are a completely different situation.  The ground state of 
three such spins interacting through antiferromagnetic Heisenberg exchange 
is any state with a total spin $S_\triangle = 1/2$. Such a state is fourfold
degenerate: part of this degeneracy is of the Kramers type, as the projection
of the total spin on an arbitrary axis can be $S_\triangle^z = \pm 1/2$,
and there is an additional twofold degeneracy related to the symmetry
of the triangle.  This degeneracy can be understood in terms of valence-bond
states, in which two of the spins on the triangle form a singlet bond
(a quantum dimer) while the third remains free.  Figure \ref{fig:vbs-triangle}
shows three such states, although they are not mutually orthogonal; in fact 
only two of these states are linearly independent.

The presence of a non-Kramers degeneracy leads to the spin-Jahn-Teller
effect on a triangle with spins $S = 1/2$. The analysis is similar to that 
for four $S = 1/2$ spins on a tetrahedron (Sec.~\ref{sec:JT-half}), with 
three distinct ground states. Depending on the sign of the cubic term, the 
distorted triangle will have either two long bonds and one short bond with 
a singlet on it, or two short bonds and one long bond with a triplet.

\section{Models with local phonon modes}
\label{sec:local-phonon}

The symmetry-based analysis of the previous section is basically
exact regardless of the underlying microscopic model for the phonons.
In this section we review some specific models of spin-lattice
coupling based on local phonon modes and also discuss their
applications.

Probably the simplest situation is the ``bond-phonon model,'' in which 
the exchange integral (\ref{eq:H1}) and also the elastic energy depend 
only on the bond length $r_{ij}$. The elastic energy is approximated by 
the sum of individual terms $k \delta r_{ij}^2/2$. After integrating out 
the bond variables $\delta r_{ij}$, the model generates a biquadratic 
spin Hamiltonian
\begin{equation}
\label{eq:biquad}
-(J'^2/2k) \sum_{\langle ij \rangle} (\mathbf S_i \cdot \mathbf S_j)^2,
\end{equation}
which clearly favors collinear spin configurations. Because of its
simplicity, this model has been applied in numerous studies of frustrated 
magnets. As one example, Becca \textit{et al.} found using the bond-phonon 
model that magnetoelastic coupling leads to a spin-Peierls transition in 
the frustrated $J_1$-$J_2$ antiferromagnet on the square lattice 
\cite{becca:02prl,weber:05prb}. This study may be relevant to the nature 
of the transition to a phase of collinear order observed in the quasi-2D
antiferromagnet Li$_2$VOSiO$_4$.

On the pyrochlore lattice, the bond-phonon model gives a ground state with 
$3^N$-fold degeneracy, where $N$ is the total number of tetrahedra: each 
tetrahedron can be flattened along one of the 3 major axes, independently 
of the other tetrahedra. A more realistic phonon model can be formulated 
in terms of the independent displacements of each atom, with the bond 
lengths determined from these atomic displacements by $\delta r_{ij} = 
(\mathbf u_i - \mathbf u_j) \cdot \hat\mathbf r_{ij}$. This is known as 
the site-phonon (or Einstein phonon) model of spin-lattice coupling 
\cite{jia:05prb,bergman:134409}, in the simplest version of which the 
elastic energy is approximated by a sum of individual terms $k|\mathbf 
u_i|^2/2$, an assumption which leads to a constant dispersion similar 
to the long-wavelength limit of optical phonons. In addition to 
Eq.~(\ref{eq:biquad}), after integrating out the displacements the model 
introduces couplings between bond variables,
\begin{equation}
\label{eq:bond-coupling}
-(J'^2/2k) \sum_{i,\, j\neq k} 
(\hat\mathbf r_{ij}\cdot\hat\mathbf r_{ik})\,(\mathbf S_i\cdot\mathbf
S_j)\,(\mathbf S_i\cdot\mathbf S_k).
\end{equation}
Because of these coupling terms, coherent long-range distortions are 
possible in this model. Using the site-phonon model, Wang and Vishwanath
\cite{wang:08prl} showed that a zigzag collinear order of the triangular 
antiferromagnet CuFeO$_2$ is a ground state of the spin-lattice Hamiltonian. 
However, because the Fe$^{3+}$ ion in this compound has spin $S = 5/2$ and 
$L = 0$, resulting in a rather small magnetic anisotropy, a relatively 
large spin-lattice coupling is required to induce the zigzag collinear 
order from the non-collinear 120$^\circ$ ground state of Heisenberg spins 
on this lattice.

In the context of the pyrochlore lattice, the interaction term of 
Eq.~(\ref{eq:bond-coupling}) corresponds to an antiferromagnetic coupling
between the force doublets of nearest-neighbor tetrahedra, $K\sum_{\langle 
\alpha\beta \rangle} \mathbf f_{\alpha} \cdot \mathbf f_{\beta}$, where the 
coupling constant $K > 0$. As a result, neighboring tetrahedra tend to be 
flattened along different directions (\textit{e.g.}~$\langle 100 \rangle$ 
and $\langle 010 \rangle$). While this reduces the total number of ground 
states, it still leaves a very high accidental degeneracy which grows 
exponentially with $N$.

\subsection{Half-magnetization plateau in $A$Cr$_2$O$_4$ spinels}
\label{sec:plateau}

Local phonon models provide an explicit description with which one
may study the spin-lattice stabilization of the half-magnetization
plateau observed in some spinel chromites. The low-temperature
magnetization curves of the spinels CdCr$_2$O$_4$ and HgCr$_2$O$_4$
exhibit a sharp transition into a wide plateau region where the 
magnetization is equal to half its saturation value \cite{ueda:05prl,
ueda:06prb}. Each tetrahedron in the half-plateau state is in one of
the four 3-up-1-down collinear spin configurations. While thermal and 
quantum fluctuations act in general to favor collinear spins, and 
indeed in some cases help to stabilize the magnetization plateaus in
frustrated magnets, the observed half-magnetization plateau in spinels 
arises most likely due to a coupling between the spins and the lattice,
a scenario also supported by the observation of a spontaneous
structural distortion accompanying the half-plateau transition.

Because the total spin $S_{\rm tot}$ is nonzero in the presence of a
magnetic field, coupling of the moments to the singlet ($A$) and triplet
($T_2$) phonon modes can no longer be neglected. Still, when the applied 
magnetic field is weak, the distortion is such that the crystal retains a 
tetragonal symmetry. The spins develop a canted antiferromagnetic
order with two frustrated bonds and four antiferromagnetic
bonds. As the field strength increases, the doublet force $|\mathbf
f^{E}| = (4/\sqrt{3})\,(S^2 - S_{\rm tot}^2/16)$ decreases as a result
of the increasing total spin. At a critical field, the trigonal
distortion, accompanying a 3-up-1-down collinear spin configuration
maximizing the triplet forces, $f^{T_2}_i = \sqrt{2}S^2$, becomes
more favorable energetically. The tetragonal distortion thus gives
way to the trigonal one through a discontinuous transition.

Using the bond-phonon model, Penc and co-workers \cite{penc:197203}
obtained a classical phase diagram for the spin-Jahn-Teller effect 
in a single tetrahedron. The transition to the collinear 3:1 states
takes place at a field strength $H\approx 3J$ and $J'^2/k \gtrsim
0.05 J$. A general symmetry analysis also showed that the collinear
3:1 states are always stabilized over a finite range of magnetic
field, provided that $J^{E'} < 2 J^{T_2 '}$ \cite{penc:197203}.

Similar to the ground-state manifold at zero field, the model with
independent bonds retains an extensively degenerate manifold of
half-magnetization states, because the trigonal distortion of individual
tetrahedra can be along any of the four $\langle 111 \rangle$ axes.
This accidental degeneracy is lifted by the additional coupling term
(\ref{eq:bond-coupling}) introduced by the site-phonon model, which
favors an antidistortive coupling between neighboring tetrahedra.
Using this rule, Bergman \textit{et al.} \cite{bergman:134409}
showed that a unique spin configuration (up to discrete lattice
symmetries) with a quadrupled, cubic unit cell is the ground state of
the system in the half-magnetization plateau. The resulting space group, 
$P4_3 32$, is consistent with the X-ray diffraction pattern of the spinel
HgCr$_2$O$_4$ \cite{matsuda:nature-phys}.

\section{Collective Spin-Jahn-Teller effect on the pyrochlore lattice}
\label{sec:collective}

Attempts to extend the analysis of the spin-Jahn-Teller effect on
a single tetrahedron to an infinite pyrochlore lattice encounter a
conceptual problem: there are \textit{infinitely many} phonon modes
coupled to the spins (one may expect two for every tetrahedron).
There are also technical difficulties: detailed knowledge of the 
crystal's elastic properties is required. As a result of these 
difficulties, the problem lacks a general solution.

Some progress may, however, be made through the use of local phonon 
models, as described in the previous section. Still, a massive accidental
degeneracy remains in the ground states of these simplified models.
Further, the lattice modes in real crystals are plane waves, and thus a 
lattice distortion involves only a small number of phonons with specific 
lattice momenta. For example, the distortion in ZnCr$_2$O$_4$ shows 
superlattice peaks in a diffraction experiment with wavenumbers
$[\frac{1}{2}\frac{1}{2}\frac{1}{2}]$ \cite{ueda:094415}. This does 
make it possible to take an alternative, phenomenological approach in 
which only a small number of lattice modes is considered.  Such an 
approach was taken by Tchernyshyov \textit{et al.}~\cite{Tch02PRB}, 
who considered the simplest case where spin displacements preserve
the translational symmetries of the lattice and break only point-group
symmetries.

\begin{figure}
\begin{center}
\includegraphics[width=0.6\columnwidth]{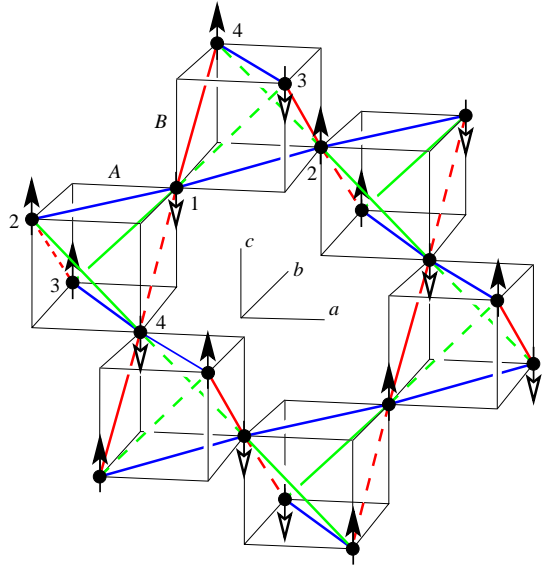}
\end{center}
\caption{Magnetic order in a state with a softer $E_u$ mode. Frustrated
bonds are shown as dashed lines. Tetrahedra of the two different
orientations (labeled $A$ and $B$) are flattened along axes $a$ and $b$, 
so that the net distortion of the lattice is (for $J' < 0$) an elongation 
along axis $c$. Reprinted with permission from Ref.~\cite{chern:060405}.}
\label{fig:Neel-u}
\end{figure}

The pyrochlore lattice is made up of tetrahedra of two different orientations.
Because all tetrahedra of the same orientation are related by lattice 
translation (which is assumed to remain a good symmetry), it is necessary 
to consider only two tetrahedra of opposite orientations $A$ and $B$ 
(Fig.~\ref{fig:Neel-u}). The symmetry group must be expanded from $T_d$ by 
including the inversion $I$ exchanging the two sublattices of tetrahedra, 
$T_d \otimes I = O_h$ \cite{LL}. The irreps remain largely unaltered, with 
the exception of a newly added parity index, which enters because these are 
either even ($g$) or odd ($u$) under the inversion.

At linear order in the displacements, the only modes which couple to the 
spins are the doublets $E_g$ and $E_u$. The former represents an overall
tetragonal or orthorhombic distortion of the lattice, while the latter is
an optical phonon with wavenumber $\mathbf q = 0$ that distorts tetrahedra 
of opposite orientations in exactly opposite ways (\textit{e.g.}~by 
flattening tetrahedra $A$ and elongating tetrahedra $B$ by the same 
amount and in the same direction). These modes can be expressed in terms 
of linear combinations of distortions on tetrahedra of types $A$ and $B$,
\begin{equation}
\mathbf Q^g = \frac{\mathbf Q^A + \mathbf Q^B}{\sqrt{2}}, \quad
\mathbf Q^u = \frac{\mathbf Q^A - \mathbf Q^B}{\sqrt{2}}.
\end{equation}
The spin-lattice energy (\ref{eq:H}) generalizes to
\begin{equation}
E(\mathbf f^A, \mathbf f^B, \mathbf Q^A, \mathbf Q^B)
  = J'(\mathbf Q^A \cdot \mathbf f^A + \mathbf Q^B \cdot \mathbf f^B)
  + k_g|\mathbf Q^g|^2/2 + k_u|\mathbf Q^u|^2/2,
\label{eq:H-AB}
\end{equation}
where $k_g$ and $k_u$ are the elastic constants of the even and odd $E$ 
doublets. Minimization with respect to the lattice modes $\mathbf Q^g$ 
and $\mathbf Q^u$ yields the energy as a function of the spin variables 
in the form 
\begin{eqnarray}
E(\mathbf f^A, \mathbf f^B)
& = & -\frac{K_g|\mathbf f^A + \mathbf f^B|^2}{4}
  -\frac{K_u|\mathbf f^A - \mathbf f^B|^2}{4}
\nonumber\\
& = & -\frac{\left(K_g + K_u\right)
         \left(|\mathbf f^A|^2 + |\mathbf f^B|^2 \right)}{4}
\label{eq:E-f}\\
  &&-\frac{\left(K_g - K_u\right)
         \left(\mathbf f^A \cdot \mathbf f^B \right)}{2},
\nonumber
\end{eqnarray}
where we have introduced the effective magnetoelastic exchange
couplings $K_{g,u} = {J'}^2/k_{g,u}$.

The second line in Eq.~(\ref{eq:E-f}) is the result familiar from 
Eq.~(\ref{eq:E-cl}): the magnitude of the doublet force $\mathbf f$ is 
maximized on tetrahedra of both sublattices. Thus one expects to find 
states with collinear spins and (for $J' < 0$) tetrahedra flattened along 
one of the three $\langle 100 \rangle$ directions.  The third line in 
Eq.~(\ref{eq:E-f}) represents a coupling between the $\mathbf f$ variables 
of the two sublattices, whose consequences depend on which of the two 
lattice modes is softer.

\begin{figure}
\begin{center}
\includegraphics[width=0.6\columnwidth]{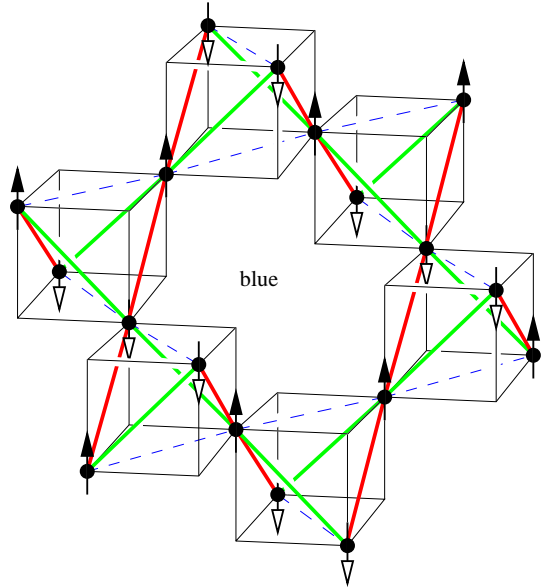}
\end{center}
\caption{Magnetic order in a state with a softer $E_g$ mode. Frustrated
bonds are shown as dashed lines. The lattice is flattened uniformly 
(for $J' < 0$).  Reprinted with permission from Ref.~\cite{Tch02PRB}.}
\label{fig:Neel-g}
\end{figure}

If $K_g > K_u$ ($E_g$ mode softer), the energy is minimized when
$\mathbf f^A$ and $\mathbf f^B$ are in the same corner of the force
triangle (Fig.~\ref{fig:f}). Tetrahedra of both sublattices are flattened  
along the same $\langle 100 \rangle$ direction, so that only an $E_g$ 
distortion is present. The spin configuration is shown in 
Fig.~\ref{fig:Neel-g}. The magnetic unit cell coincides with the 
structural one. Because we are considering an O(3)-symmetric Heisenberg 
model, the global orientation of the spins can be arbitrary 

If, on the other hand, $K_g < K_u$ (softer $E_u$ mode), the two $\mathbf f$
vectors are located in different corners of the triangle, so that tetrahedra 
of types $A$ and $B$ are flattened along two different $\langle 100 \rangle$ 
directions, giving six possible ground states. The distortion is a
superposition of the $E_u$ and $E_g$ modes. The presence of the even mode 
is understood in a straightforward manner: if tetrahedra of type $A$ are 
flattened along $\langle 100 \rangle$ and tetrahedra of type $B$ along 
$\langle 010 \rangle$, the lattice is on average \textit{elongated} along 
$\langle 001 \rangle$. The presence of the $E_u$ component of the distortion 
means that the inversion symmetry is broken spontaneously. This has important 
consequences for the magnetic order, shown in Fig.~\ref{fig:Neel-u}, as we 
discuss in detail below. Here we note only that frustrated bonds form 
left-handed spirals in this particular state.

\section{Collective Jahn-Teller effect in CdCr$_2$O$_4$}
\label{sec:cdcr2o4}

The normal spinels $A$Cr$_2$O$_4$ (where $A$ is a nonmagnetic Zn, Mg, 
Cd, or Hg ion) are strongly frustrated antiferromagnets exhibiting the 
spin-induced Jahn-Teller distortion. The magnetic Cr$^{3+}$ ions forming 
the pyrochlore lattice have electronic configuration $3d^3$. The oxygen 
octahedron surrounding the chromium ion splits the $3d$ levels into a 
high-energy $e_g$ doublet and a low-energy $t_{2g}$ triplet; the former 
are unoccupied while the latter are singly occupied, with the 3 electrons 
having parallel spins. Thus the orbital degree of freedom is quenched and 
the spins form a nearly isotropic magnetic moment with total spin $S = 3/2$ 
and a gyromagnetic ratio of $g = 2$ \cite{PhysRevLett.84.3718}. Interactions 
between the spins are mediated mostly by direct antiferromagnetic exchange 
between neighboring Cr sites \cite{sushkov:137202} (with the exception of 
HgCr$_2$O$_4$, where this contribution is comparable to the ferromagnetic 
superexchange term mediated by the oxygen ions).

All of these compounds have a strongly correlated paramagnetic state well 
below their Curie-Weiss temperatures, $\Theta$, and order magnetically at 
a temperature $T_N \ll \Theta$. The phase transition is discontinuous and 
is accompanied by a lattice distortion \cite{PhysRevLett.84.3718, 
chung:247204, ueda:094415}. The lack of orbital degrees of freedom points 
to a magnetoelastic origin for the lattice distortion.

CdCr$_2$O$_4$ is the best-understood system to date. In the distorted state
below $T_N$ it shows no superlattice peaks \cite{chung:247204}, indicating
that the translational symmetry of the high-temperature cubic phase
(space group $Fd\bar{3}m$) remains unchanged.  The point-group symmetry
is lowered: the lattice exhibits a tetragonal distortion with lattice
constants $a = b < c$ (an overall elongation). The low-temperature structure
was identified by Chung \textit{et al.} as the pyrochlore lattice with a 
uniform tetragonal elongation (space group $I4_1/amd$) \cite{chung:247204}, 
\textit{i.e.}~a pure $E_g$ distortion.  However, as listed next, there are 
good reasons to believe that this distortion also involves the staggered 
component $E_u$, which breaks the inversion symmetry and lowers the space 
group down to $I4_122$. 

1. The dominance of direct antiferromagnetic exchange between adjacent
chromium spins means that the exchange constant decreases with increasing 
ionic separation, \textit{i.e.}~$J'<0$. If the distortion were of pure 
$E_g$ type, the crystal would flatten along one axis, yielding $a = b > c$ 
in contradiction to the experimental data \cite{chung:247204}. As explained 
in the previous section, an $E_u$-driven distortion would lead to an overall 
elongation of the lattice, $a = b < c$, in agreement with the experiment.

2. An $E_u$ distortion breaks the inversion symmetry of the crystal,
making the crystalline lattice \textit{chiral}. Indeed, the elongated
(frustrated) bonds in Fig.~\ref{fig:Neel-u} form spirals of one helicity. 
Spin-orbit coupling in the form of the Dzyaloshinskii-Moriya interaction 
would then spread the chirality from the lattice to the spins, generating 
a spiral magnetic order. The observed magnetic order in CdCr$_2$O$_4$ is 
in fact a weakly incommensurate spiral \cite{chung:247204}.

\subsection{Spiral magnetic order in CdCr$_2$O$_4$}
\label{sec:cdcr2o4-spiral}

Chung \textit{et al.}~reported an incommensurate magnetic order with
magnetic Bragg peaks at $\mathbf q = 2\pi(0, \delta, 1)$ in a crystal 
with an elongated $c$ axis ($a = b = 0.995c$) and $\delta = 0.09$. The
magnetization lies in the $ac$-plane. Because the incommensurability
$\delta$ is small, the magnetic order can be understood as a commensurate
state with $\mathbf q = 2\pi(0,0,1)$ twisting slowly along the $b$-direction.

The same authors found two such structures which would be consistent with 
the data they obtained from elastic neutron scattering. One of the proposed 
ordering patterns is derived from the commensurate state shown in 
Fig.~\ref{fig:Neel-u}, and is precisely what one expects when the 
magnetoelastic effect is driven by the $E_u$ phonon. The other candidate 
is derived from an ``orthogonal'' state where the spins on every tetrahedron 
are oriented, for example, in directions $+\hat \mathbf x$, $-\hat \mathbf 
x$, $+\hat \mathbf y$, and $-\hat \mathbf y$. Such a state is very hard to 
obtain through the spin-driven Jahn-Teller effect \cite{Tch02PRB}, and no 
other justification for this state is currently known.

The small value of $\delta$ makes it possible to treat this 
incommensurability as the result of a weak perturbation on top of 
the Heisenberg exchange and magnetoelastic coupling. Chern \textit{et 
al.}~\cite{chern:060405} derived possible magnetic spiral states that 
arise when the Dzyaloshinskii-Moriya interaction is added to these two 
energy terms. These authors found two candidate solutions, one of which 
is entirely consistent with the data of Chung \textit{et al.} This 
analysis and its results are described in the next section.

\subsection{Theory of spiral magnetic order}
\label{sec:cdcr2o4-theory}

The Dzyaloshinskii-Moriya (DM) interaction gives a contribution 
\begin{equation}
H_\mathrm{DM} = \sum_{\langle ij \rangle}
\mathbf D_{ij}\,\cdot\, [\mathbf S_i \times \mathbf S_j],
\label{eq:H-DM}
\end{equation}
to the Hamiltonian, where the sum extends over pairs of nearest neighbors. 
This term is allowed on the ideal pyrochlore lattice, where the bonds are 
not centrosymmetric, the precondition for a non-vanishing coupling constant 
$\mathbf D_{ij}$.  At the same time, the high symmetry of the lattice 
constrains the direction of this vector \cite{elhajal:094420}: for a bond 
oriented along the $[110]$ lattice direction, the vector
must point along $[1\bar{1}0]$ ($\mathbf D_{ij} = (\pm D, 
\mp D, 0)/\sqrt{2}$). The value of $\mathbf D_{ij}$ on any other bond is 
then found through the symmetry transformations of the system. In a 
collinear magnetic state, the expectation value of $H_\mathrm{DM}$ is 
zero, but its contribution can be lowered by twisting the spins into a 
spiral.

The pitch of the spiral is determined by the competition between the
DM coupling strength $D$ and a spin stiffness. In most antiferromagnets,
the spin stiffness is set by the exchange interaction, but the pyrochlore 
antiferromagnet with nearest-neighbor exchange is an exception: the large 
degeneracy of its ground state leads to a vanishing stiffness. As a result, 
the ground state in the presence of the DM interaction is not a long-range 
spiral but is rather a commensurate state with noncollinear spins 
\cite{elhajal:094420}.

The presence of magnetoelastic interactions changes the situation.
Recall that the spin-induced Jahn-Teller effect selects a state with
collinear spins. This tendency towards collinear states results in
a finite spin stiffness. As a result, the pitch of the spiral is
a quantity of order $D/K$, where $K = {J'}^2/k$ is the effective
magnetoelastic interaction.

A problem where the nearest-neighbor exchange $J$, the magnetoelastic 
energy scale $K$, and the DM coupling $D$ are each arbitrary is hard to
solve analytically. However, it simplifies if these energy scales are 
well separated, the conventional hierarchy being 
\begin{equation}
JS^2 \gg KS^4 \gg DS^2.
\label{eq:separation}
\end{equation}
A quantitative analysis indicates that in CdCr$_2$O$_4$ these scales are
of similar magnitude (as discussed at the end of this section), but still
have the expected order, $JS^2 > KS^4 > DS^2$. Thus while a theory based 
on the assumption (\ref{eq:separation}) may not provide a quantitative 
account of magnetic order in CdCr$_2$O$_4$, it presents at minimum a good 
point of departure for understanding it.

Minimization of the dominant term, the exchange energy (\ref{eq:H0}), yields
a constraint that the total spin be zero on every tetrahedron.  The remaining
degrees of freedom of a single tetrahedron can be parametrized using the 
staggered magnetizations
\begin{eqnarray*}
{\bf L}_1 = \frac{{\bf S}_1 - {\bf S}_2 - {\bf S}_3 + {\bf S}_4}{4S},
{\bf L}_2 = \frac{-{\bf S}_1 + {\bf S}_2 - {\bf S}_3 + {\bf S}_4}{4S},
{\bf L}_3 = \frac{-{\bf S}_1 - {\bf S}_2 + {\bf S}_3 + {\bf S}_4}{4S}.
\end{eqnarray*}
Because each spin belongs to two tetrahedra, the staggered magnetizations
on one sublattice of tetrahedra determine completely those of the
other sublattice. We use the staggered magnetizations of sublattice
$A$, $\{\mathbf L_i^A\}$, to express the magnetizations of sublattice 
$B$, and, except in cases of possible confusion, suppress the sublattice 
index to simplify the notation.

Even for a single tetrahedron, the three staggered magnetizations are
not completely independent. Vanishing of the total spin in the ground
state makes them mutually orthogonal and imposes on their lengths the 
constraint \cite{henley:3962}
\begin{equation}
{\bf L}_i = L_i \hat \mathbf l_i, \quad
\hat \mathbf l_i \cdot \hat \mathbf l_j = \delta_{ij}, \quad
\sum_{i=1}^3 L_i^2=1.
\label{eq:length-constraint}
\end{equation}
The lengths $L_i$ parametrize the angles between the spins and are 
related to the bond doublet ${\bf f}$ by 
\begin{eqnarray}
    f_1 = 2S^2\,(L_1^2+L_2^2-2L_3^2)/\sqrt{3},
    \quad
    f_2 = 2S^2\,(L_1^2-L_2^2).
    \label{eq:f-phi}
\end{eqnarray}
Thus five parameters are required to describe the magnetic state of a 
tetrahedron in the ground state of Eq.~(\ref{eq:H0}): three Euler angles
for the triad $\{\hat{\bf l}_i\}$ and two further parameters for the bond 
doublet, \textit{e.g.}~${\bf f}=(f_1,f_2)$.

The magnetic order of CdCr$_2$O$_4$ in the commensurate limit $\delta \to 0$
(Fig.~\ref{fig:Neel-u}) has staggered site magnetizations: ${\bf L}_2 =
{\bf L}_3 = 0$ and ${\bf L}_1 = e^{i{\bf q} \cdot {\bf r}}\,\hat{\bf n}_1$, 
where ${\bf q}= 2\pi(0,0,1)$ and $\hat \mathbf n_1$ is an arbitrary unit 
vector. In terms of the three staggered magnetizations, the DM term for a 
single tetrahedron is
\begin{eqnarray}
  E_\mathrm{DM}=-DS^2\, (\hat{\bf a}\cdot\,{\bf L}_2\times{\bf L}_3
  +\hat{\bf b}\cdot\,{\bf L}_3\times{\bf L}_1
  +\hat{\bf c}\cdot\,{\bf L}_1\times{\bf L}_2).
  \label{eq:DM2}
\end{eqnarray}
It is easy to see that the DM energy is exactly zero for the collinear
state. For either sign of $D$, this term can be made negative by allowing
a small component of ${\bf L}_2$ or ${\bf L}_3$, which describes a twisting
spin configuration. The lowering of the DM energy (\ref{eq:DM2}) is
accompanied by an increase of the magnetoelastic energy. However, further
analysis shows that the former is linear in ${\bf L}_2$ and ${\bf L}_3$
while the latter is quadratic, so that such a twisting distortion always 
occurs.

To pass to a continuum description, we express the rapidly oscillating
unit vectors $\hat \mathbf l_i = \hat \mathbf n_i \, e^{i\mathbf {q \cdot r}}$
in terms of a slowly varying orthonormal triad $\{\hat \mathbf n_i\}$,
and use the length constraint (\ref{eq:length-constraint}) to eliminate 
$L_1$ by 
\begin{equation}
    \mathbf L_1 = \left(1-\frac{L_2^2+L_3^2}{2}\right)\,
    \hat{\bf n}_1 \, e^{i\mathbf {q \cdot r}}, \quad
    \mathbf L_2 = L_2\,\hat{\bf n}_2 \, e^{i\mathbf {q \cdot r}}, \quad
    \mathbf L_3 = L_3\,\hat{\bf n}_3 \, e^{i\mathbf {q \cdot r}}.
\end{equation}
At this point, the magnetic structure is described in terms of five slowly
varying fields, $L_2$, $L_3$, and the triad $\{\hat \mathbf n_i\}$.

The number of independent fields can be further reduced by examining
tetrahedra of sublattice $B$: the vector of total magnetization on each 
$B$ tetrahedron must vanish, giving three more constraints. The total
spin of a $B$ tetrahedron centered at $\mathbf r^B = \mathbf r^A +
(1/4,1/4,1/4)$ is given by
\[
\mathbf M^B(\mathbf r^B)
    = \mathbf S_1(\mathbf r^A + \mathbf a_1)
    + \mathbf S_2(\mathbf r^A + \mathbf a_2)
    + \mathbf S_3(\mathbf r^A + \mathbf a_3)
    + \mathbf S_4(\mathbf r^A),
\]
where ${\bf a}_1 = (0,1/2,1/2)$, ${\bf a}_2 = (1/2,0,1/2)$, and ${\bf a}_3 
= (1/2,1/2,0)$ are primitive lattice vectors (the centers of the tetrahedra 
form a diamond lattice; the Bravais lattice is fcc). Expressing the spins
$\{\mathbf S_i\}$ in terms of the staggered magnetizations $\{\mathbf L_i\}$ 
and using the gradient expansion, one obtains, to the lowest orders in $L_2$, 
$L_3$, and the gradients, the constraint
\begin{equation}
  \mathbf M^B = L_3\,\hat{\bf n}_3 -
  \frac{1}{4}\frac{\partial \hat{\bf n}_1}{\partial y} = 0.
  \label{eq:phi3}
\end{equation}
Thus it is clear that $L_3$ and $\hat \mathbf n_3$ are determined by the 
gradient $\nabla \hat \mathbf n_1$.

In a similar way, expressing the staggered magnetizations on sublattice 
$B$ to lowest order in $\nabla\hat{\bf n}_1$ leads to 
\begin{eqnarray}
  \mathbf L^B_1 = L_2\hat{\bf n}_2-\frac{1}{4}
  \frac{\partial \hat{\bf n}_1}{\partial z}, \quad
  \mathbf L^B_2=\hat{\bf n}_1, \quad
  \mathbf L^B_3 = -\frac{1}{4}\frac{\partial \hat{\bf n}_1}{\partial x}.
  \label{eq:LB}
\end{eqnarray}
Substituting these expressions into Eq.~(\ref{eq:DM2}) and adding
contributions from tetrahedra of both types yields
\begin{eqnarray}
  E_\mathrm{DM} = -\frac{1}{4}\, DS^2\, \hat{\bf n}_1 \cdot
  \Bigl(
  \hat{\bf a} \times \frac{\partial \hat{\bf n}_1}{\partial x}
  +\hat{\bf b} \times \frac{\partial \hat{\bf n}_1}{\partial y}
  -\hat{\bf c} \times \frac{\partial \hat{\bf n}_1}{\partial z}
  \Bigr).
  \label{eq:EDM}
\end{eqnarray}
The DM energy contains terms linear in gradients of $\hat{\bf n}_1$,
indicating that the spins are unstable against the formation of spiral 
configurations. As an example, the first term $\hat{\bf n}_1 \cdot 
\hat{\bf a} \times \partial\hat{\bf n}_1/\partial x$ favors a magnetic 
order with the unit vector $\hat{\bf n}_1$ perpendicular to the $a$-axis 
and spiraling about it.

As discussed previously, the spin stiffness has its origin in the 
magnetoelastic energy. For simplicity, here we consider only distortions 
due to $E_u$ phonons and neglect the effect of $E_g$ distortions (a procedure 
definitely appropriate in the limit $K_g \ll K_u$). The linear decrease of 
the DM energy due to gradients of $\hat{\bf n}_1$ must be balanced by the 
increase in magnetoelastic energy, which on symmetry grounds must be 
quadratic in $\nabla \hat{\bf n}_1$. From Eq.~(\ref{eq:E-f}), the increase 
of magnetoelastic energy is $E_{\rm me} =  -K_u \, {\bf u}_0 \cdot 
\delta{\bf u}/2$, where $K_u = J'^2/k_u$ and ${\bf u} = {\bf f}^A - {\bf 
f}^B$. The unperturbed odd doublet is ${\bf u}_0 = 4S^2\,(0,1)$. Because 
we are describing the spiral magnetic order in terms of the staggered 
magnetizations, it is convenient to use Eq.~(\ref{eq:f-phi}) for the 
calculation of $\delta{\bf f}^A$ and $\delta{\bf f}^B$. Retaining terms 
to second order in $\nabla\hat{\bf n}_1$ leads to 
\begin{eqnarray}
    E_{\rm me} = \frac{K_u S^4}{4} \! \left[
    \left(\frac{\partial\hat{\bf n}_1}{\partial x}\right)^2 \!\!
    + \! \left(\frac{\partial\hat{\bf n}_1}{\partial y}\right)^2 \!\!
    + \! 2 \! \left(\frac{\partial\hat{\bf n}_1}{\partial z}\right)^2 \!\!
    - \! L_2\hat{\bf n}_2\cdot\frac{\partial\hat{\bf n}_1}{\partial z}
    + 4L_2^2 \right] \! .
    \label{eq:Eme1}
\end{eqnarray}
Because the DM energy (\ref{eq:EDM}) does not depend on $L_2$, minimization
of the total energy with respect to $L_2$ affects only the magnetoelastic 
term (\ref{eq:Eme1}) and yields
\begin{equation}
    L_2 = \frac{1}{8}\,\hat{\mathbf n}_2 \cdot
    \frac{\partial \hat{\mathbf n}_1}{\partial z};
\label{eq:phi2}
\end{equation}
thus $L_2$ is also eliminated. The minimized magnetoelastic energy is
\begin{eqnarray}
  E_\mathrm{me} =\frac{K_u S^4}{4} \! \left[
    \left(\frac{\partial\hat{\bf n}_1}{\partial x}\right)^2 \!\!
    + \! \left(\frac{\partial\hat{\bf n}_1}{\partial y}\right)^2 \!\!
    + 2 \! \left(\frac{\partial\hat{\bf n}_1}{\partial z}\right)^2 \!\!
    - \! \left(\hat{\bf n}_2\cdot\frac{\partial\hat{\bf n}_1}{\partial 
    z}\right)^2 \right] \! .
  \label{eq:Eme2}
\end{eqnarray}

\begin{figure}
\begin{center}
\includegraphics[width=0.9\columnwidth]{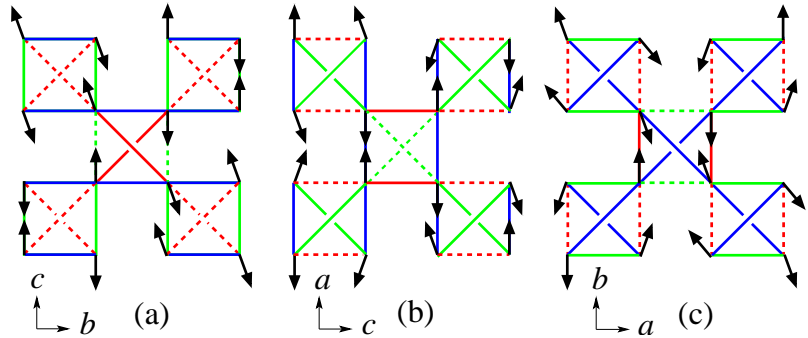}
\end{center}
\caption{Three symmetrical spiral magnetic structures minimizing the
energy. The spins are perpendicular to and twist about the $a$-axis (a), 
$b$-axis (b), or $c$-axis (c). Dashed lines indicate frustrated bonds. 
The crystal is viewed along a $\langle 100 \rangle$ direction.  
Reprinted with permission from Ref.~\cite{chern:060405}.}
\label{fig:spiral}
\end{figure}

The total energy of the spiral state, now expressed as a functional of the 
vector fields $\hat{\bf n}_1({\bf r})$ and $\hat{\bf n}_2({\bf r})$, is the 
sum of Eqs.~(\ref{eq:EDM}) and (\ref{eq:Eme2}). Its minimization yields
a second-order partial differential equation. While we are unable to find 
the most general solution to this equation, we can find three highly
symmetrical spiral solutions in which the spins remain perpendicular to, 
while twisting about, one of the $\langle 100 \rangle$ axes. As one example, 
a spiral state along the $b$ axis is described by $\hat{\bf n}_1 = (\cos 
\theta(y),0,\sin\theta(y))$ and has total energy
\begin{equation}
\label{eq:Etot}
E = -(DS^2 \theta' + K_u S^4 {\theta'}^2)/4,
\end{equation}
where $\theta' = d\theta/d y$. Minimization of this quantity gives the 
pitch of the spiral,
\begin{equation}
 \theta' = 2\pi\delta =  \frac{D}{2 K_u S^2}.
 \label{eq:delta}
\end{equation}
Equation (\ref{eq:phi2}) implies that ${\bf L}^A_2 = 0$; the $A$ tetrahedra 
therefore have coplanar spins spanned by two orthogonal N\'eel vectors, 
${\bf L}^A_1$ and ${\bf L}^A_3$. The angles between spin pairs are given 
by $\theta_{14} =\theta_{23} = 2L_3$. From Eq.~(\ref{eq:phi3}), this angle 
is related to the pitch by $2L_3 = \pi \delta$. The spiral magnetic state 
has the structure
\begin{eqnarray}
    {\bf L}_1^A &=& \cos {(\pi\delta/2)}\,
              \left(\cos{(2\pi\delta y)},\,0,\,\sin{(2\pi\delta y)}\right)\,
    e^{2\pi i z}, \nonumber \\
    \mathbf L_2^A &=& 0, \label{eq:spiral-b}\\
    {\bf L}_3^A &=& \sin {(\pi\delta/2)}\,
              \left(-\sin{(2\pi\delta y)},\,0,\,\cos{(2\pi\delta y)}\right)\,
    e^{2\pi i z},
    \nonumber
\end{eqnarray}
producing Bragg scattering at wavevector ${\bf q} = 2\pi (0,\pm 
\delta, 1)$, while the ordered magnetic moments lie in the $ac$-plane 
[Fig.~\ref{fig:spiral}(b)]. All of this is consistent with the experimental 
data of Ref.~\cite{chung:247204}.

It is worth noting that the distorted crystal structure preserves certain 
symmetries on interchanging the $A$ and $B$ sublattices, such as inversion 
at a Cr site followed by a $\pi/2$ rotation in the $ab$-plane. However, 
the magnetic order described here breaks these symmetries. From 
Eq.~(\ref{eq:LB}) one has ${\bf L}^B_1 = {\bf L}^B_3 = 0$, which means
that every tetrahedron on sublattice $B$ has collinear spins, whereas the 
spins of the $A$ tetrahedra are twisted into a (weakly) non-collinear state. 
This disparity between the two sublattices should result in different
distortions of the two types of tetrahedra, thus further lowering the
symmetry of the crystal. However, the magnitude of the additional 
distortion is expected to be small because the degree of non-collinearity 
is small, $\delta \ll 1$.

Spiral states in which the spins rotate about the $a$-axis 
[Fig.~\ref{fig:spiral}(a)] can be obtained similarly, by using the 
ansatz $\hat{\bf n}_1 = (0,\cos\theta(x),\sin\theta(x))$. The resulting 
solution can also be obtained from Eq.~(\ref{eq:spiral-b}) through symmetry 
operations which exchange the two sublattices of tetrahedra, such as 
inversion on a Cr site. This spiral produces a magnetic Bragg peak at 
${\bf q} = (\mp\delta, 0, 1)$.

Finally, there is a third spiral solution, shown in Fig.~\ref{fig:spiral}(c),
where the spins twist about the $c$-axis. In this state, which is not related 
to the other two solutions by any symmetry, the magnetic Bragg peak occurs at 
${\bf q} = 2\pi (0,0,1+\delta)$ and both sublattices have tetrahedra with 
coplanar, rather than collinear, spins. That this spiral state has the same
energy as the previous two is a coincidence: its total energy is also given 
by Eq.~(\ref{eq:Etot}) with $\theta' = d\theta/d z$. This degeneracy is 
lifted when other perturbations, such as further-neighbor interactions,
are taken into account. CdCr$_2$O$_4$ has a significant 
\textit{third}-neighbor antiferromagnetic exchange interaction, which 
acts to favor strongly the states with spirals twisting along the $a$- or 
$b$-axis \cite{chern:060405}.

In closing this section, we comment on the assumption (\ref{eq:separation}) 
of well-separated energy scales in the problem. \textit{Ab initio} 
calculations \cite{chern:060405} yield a nearest-neighbor exchange energy 
$J S^2 = 1.1$ meV and a magnetoelastic energy $K_u S^4 = 0.76$ meV. The 
strength of the DM interaction can be estimated from the measured pitch 
of the spiral, $\delta = 0.09$, using Eq.~(\ref{eq:delta}), which gives 
$DS^2 = 0.21$ meV. While the three energy scales are not vastly different, 
they do appear in the correct order of decreasing magnitude, $JS^2 > K_u 
S^4 > DS^2$.

\section{Summary and open questions}
\label{sec:questions}

The spinel compound CdCr$_2$O$_4$ provides an opportunity to test
our understanding of the ground state of the Heisenberg antiferromagnet
on the pyrochlore lattice. When the lattice degrees of freedom are included, 
both the selected magnetic order and the lattice distortion are in agreement 
with a theoretical model \cite{Tch02PRB} based on two vibrational doublets 
of the crystal, the $\mathbf q = 0$ optical phonon $E_u$ and the uniform 
lattice distortion $E_g$. The model ties the incommensurate nature of the 
spiral magnetic order to a spontaneous breaking of the inversion symmetry 
in the crystal, which has not yet been observed directly in CdCr$_2$O$_4$.

The magnetoelastic phase transition between the high-$T$ correlated
paramagnet \cite{shlee:nature} and the low-$T$ ordered phase remains
poorly understood. It is strongly discontinuous in both ZnCr$_2$O$_4$
\cite{PhysRevLett.84.3718} and CdCr$_2$O$_4$ \cite{chung:247204}, with
both the lattice distortion and the ordered moment reaching their $T = 0$ 
values immediately below the ordering temperature. A Landau free-energy
approach based on the spin-Peierls order parameter (\ref{eq:f}) appears to 
be the only candidate approach available for modeling the underlying physics 
\cite{Tch02PRB}. However, it is not evident that this phenomenology can 
provide a full description. In CdCr$_2$O$_4$, the order parameter has the 
$E_u$ symmetry and does not allow for a cubic invariant in the free energy, 
thus excluding the most obvious cause for a discontinuous phase transition. 
The phase transition may yet turn out to be first-order if the even 
($E_g$-symmetric) order parameter is nearly soft \cite{Tch02PRB}, but 
this question has not yet been clarified. In fact it remains unclear 
whether the valence-bond variables (\ref{eq:f}) represent a good choice 
of the order parameter for these systems: the low-$T$ phase in both 
ZnCr$_2$O$_4$ and CdCr$_2$O$_4$ is magnetically ordered, suggesting that 
a type of staggered magnetization may be a more appropriate choice.

Finally, a realistic model of this phase transition must take into account
the entropy of the correlated paramagnetic state \cite{Moessner98PRB}.
The very high entropy of the disordered phase may be responsible for the
discontinuous nature of the phase transition, as has been demonstrated
for the case of lattice models with large numbers of flavors 
\cite{Shlosman81}.

\bibliography{frustratedmagnets}

\begin{thebibliography}{10}
\providecommand{\url}[1]{{#1}}
\providecommand{\urlprefix}{URL }
\expandafter\ifx\csname urlstyle\endcsname\relax
  \providecommand{\doi}[1]{DOI \discretionary{}{}{}#1}\else
  \providecommand{\doi}{DOI \discretionary{}{}{}\begingroup
  \urlstyle{rm}\Url}\fi

\bibitem{Ramirez91JAP}
A.P. Ramirez, J. Appl. Phys. \textbf{70}, 5952 (1991)

\bibitem{Huse92PRB}
D.A. Huse, A.D. Rutenberg, Phys. Rev. B \textbf{45}, 7536 (1992)

\bibitem{Zinkin94PRL}
M.J. Harris, M.P. Zinkin, Z.~Tun, B.M. Wanklyn, I.P. Swainson, Phys. Rev. Lett.
  \textbf{73}, 189 (1994)

\bibitem{Moessner98PRB}
R.~Moessner, J.T. Chalker, Phys. Rev. B \textbf{58}, 12049 (1998)

\bibitem{Chalker92PRL}
J.T. Chalker, P.C.W. Holdsworth, E.F. Shender, Phys. Rev. Lett. \textbf{68},
  855 (1992)

\bibitem{Kittel60}
C.~Kittel, Phys. Rev. \textbf{120}, 335 (1960)

\bibitem{PhysRevB.14.3036}
I.S. Jacobs, J.W. Bray, H.R. Hart, L.V. Interrante, J.S. Kasper, G.D. Watkins,
  D.E. Prober, J.C. Bonner, Phys. Rev. B \textbf{14}, 3036 (1976)

\bibitem{PhysRevLett.84.3718}
S.H. Lee, C.~Broholm, T.H. Kim, W.~Ratcliff, S.W. Cheong, Phys. Rev. Lett.
  \textbf{84}, 3718 (2000)

\bibitem{chung:247204}
J.H. Chung, M.~Matsuda, S.H. Lee, K.~Kakurai, H.~Ueda, T.J. Sato, H.~Takagi,
  K.P. Hong, S.~Park, Phys. Rev. Lett. \textbf{95}, 247204 (2005)

\bibitem{ueda:094415}
H.~Ueda, H.~Mitamura, T.~Goto, Y.~Ueda, Phys. Rev. B \textbf{73}, 094415 (2006)

\bibitem{Bersuker}
I.~Bersuker, \emph{The Jahn-Teller effect} (Cambridge University Press,
  Cambridge, 2006)

\bibitem{Jahn37}
H.A. Jahn, E.~Teller, Proc. Roy. Soc. London Ser. A \textbf{161}, 220 (1937)

\bibitem{LL}
L.D. Landau, E.M. Lifshitz, \emph{Quantum mechanics: non-relativistic theory}
  (Butterworth-Heinemann, 1981)

\bibitem{PhysRevLett.85.4960}
Y.~Yamashita, K.~Ueda, Phys. Rev. Lett. \textbf{85}, 4960 (2000)

\bibitem{Tch02PRB}
O.~Tchernyshyov, R.~Moessner, S.L. Sondhi, Phys. Rev. B \textbf{66}, 064403
  (2002)

\bibitem{harris:5200}
A.B. Harris, A.J. Berlinsky, C.~Bruder, J. Appl. Phys. \textbf{69}, 5200 (1991)

\bibitem{motida:1970}
K.~Motida, S.~Miyahara, J. Phys. Soc. Jpn. \textbf{28}, 1188 (1970)

\bibitem{sushkov:137202}
A.B. Sushkov, O.~Tchernyshyov, W.~{Ratcliff II}, S.W. Cheong, H.D. Drew, Phys.
  Rev. Lett. \textbf{94}, 137202 (2005)

\bibitem{aguilar:092412}
R.V. Aguilar, A.B. Sushkov, Y.J. Choi, S.W. Cheong, H.D. Drew, Phys. Rev. B
  \textbf{77}, 092412 (2008).
\newblock \doi{10.1103/PhysRevB.77.092412}

\bibitem{olariu:167203}
A.~Olariu, P.~Mendels, F.~Bert, B.G. Ueland, P.~Schiffer, R.F. Berger, R.J.
  Cava, Phys. Rev. Lett. \textbf{97}(16), 167203 (2006)

\bibitem{PhysRevLett.99.137207}
Y.~Okamoto, M.~Nohara, H.~Aruga-Katori, H.~Takagi, Phys. Rev. Lett.
  \textbf{99}(13), 137207 (2007)

\bibitem{wang:08prl}
F.~Wang, A.~Vishwanath, Phys. Rev. Lett. \textbf{100}, 077201 (2008)

\bibitem{rudolf:024421}
T.~Rudolf, C.~Kant, F.~Mayr, A.~Loidl, Phys. Rev. B \textbf{77}, 024421 (2008).
\newblock \doi{10.1103/PhysRevB.77.024421}

\bibitem{becca:02prl}
F.~Becca, F.~Mila, Phys. Rev. Lett. \textbf{89}, 037204 (2002)

\bibitem{weber:05prb}
C.~Weber, F.~Becca, F.~Mila, Phys. Rev. B \textbf{72}, 024449 (2005)

\bibitem{jia:05prb}
C.~Jia, J.H. Nam, J.S. Kim, J.H. Han, Phys. Rev. B \textbf{71}, 212406 (2005)

\bibitem{bergman:134409}
D.L. Bergman, R.~Shindou, G.A. Fiete, L.~Balents, Phys. Rev. B \textbf{74},
  134409 (2006)

\bibitem{ueda:05prl}
H.~Ueda, H.~Katori, H.~Mitamura, T.~Goto, H.~Takagi, Phys. Rev. Lett.
  \textbf{94}, 047202 (2005)

\bibitem{ueda:06prb}
H.~Ueda, H.~Mitamura, T.~Goto, Y.~Ueda, Phys. Rev. B \textbf{73}, 094415 (2006)

\bibitem{penc:197203}
K.~Penc, N.~Shannon, H.~Shiba, Phys. Rev. Lett. \textbf{93}, 197203 (2004)

\bibitem{matsuda:nature-phys}
M.~Matsuda, H.~Ueda, A.~Kikkawa, Y.~Tanaka, K.~Katsumata, Y.~Narumi, T.~Inami,
  Y.~Ueda, S.H. Lee, Nature Physics \textbf{3}, 397  (2007)

\bibitem{chern:060405}
G.W. Chern, C.J. Fennie, O.~Tchernyshyov, Phys. Rev. B \textbf{74}, 060405
  (2006)

\bibitem{elhajal:094420}
M.~Elhajal, B.~Canals, R.~Sunyer, C.~Lacroix, Phys. Rev. B \textbf{71}, 094420
  (2005)

\bibitem{henley:3962}
C.L. Henley, J. Appl. Phys. \textbf{61}, 3962 (1987)

\bibitem{shlee:nature}
S.H. Lee, C.~Broholm, W.~Ratcliff, G.~Gasparovic, Q.~Huang, T.H. Kim, S.W.
  Cheong, Nature \textbf{418}, 856 (2002)

\bibitem{Shlosman81}
S.B. Shlosman, R.~Koteck\'y, Comm. Math. Phys. \textbf{83}, 493 (1981)

\end{thebibliography}

\end{document}